\documentclass[journal,twoside,web]{ieeecolor}
\usepackage{jsen}
\usepackage{cite}
\usepackage{amsmath,amssymb,amsfonts}
\usepackage{algorithmic}
\usepackage{graphicx}
\usepackage{textcomp}
\usepackage{wrapfig}
\usepackage{booktabs,lipsum}
\usepackage{xcolor}
\usepackage[colorlinks]{hyperref}
\usepackage{listings}

\hypersetup{
    colorlinks=true,     
    linkcolor=blue,
    citecolor=green,
}

\def\BibTeX{{\rm B\kern-.05em{\sc i\kern-.025em b}\kern-.08em
    T\kern-.1667em\lower.7ex\hbox{E}\kern-.125emX}}
\definecolor{abstractbg}{rgb}{0.89804,0.94510,0.83137}
\begin{document}
\title{A lightning monitoring system for studying transient phenomena in cosmic ray observatories}
\author{J. Pe\~na-Rodr\'iguez, P. Salgado-Meza, L. Fl\'orez-Villegas and L. A. N\'u\~nez
\thanks{J. Pe\~na-Rodr\'iguez is with the Facultad de Mecatrónica, Universidad Santo Tomás, Bucaramanga-Colombia (e-mail: jesus.pena@correo.uis.edu.co). }
\thanks{L. A. N\'u\~nez is with the Escuela de F\'isica, Universidad Industrial de Santander, Bucaramanga-Colombia (e-mail: lnunez@uis.edu.co). }
\thanks{P. Salgado-Meza and L. Fl\'orez-Villegas were students of the Escuela de Ingenierías El\'ectrica, Electr\'onica y de Telecomunicaciones, Universidad Industrial de Santander, Bucaramanga-Colombia. (e-mail: pedro.salgado@correo.uis.edu.co and leonardo.florez@correo.uis.edu.co).}}

\IEEEtitleabstractindextext{%
\fcolorbox{abstractbg}{abstractbg}{%
\begin{minipage}{\textwidth}%

\begin{wrapfigure}[12]{r}{2.8in}%
\includegraphics[width=3in]{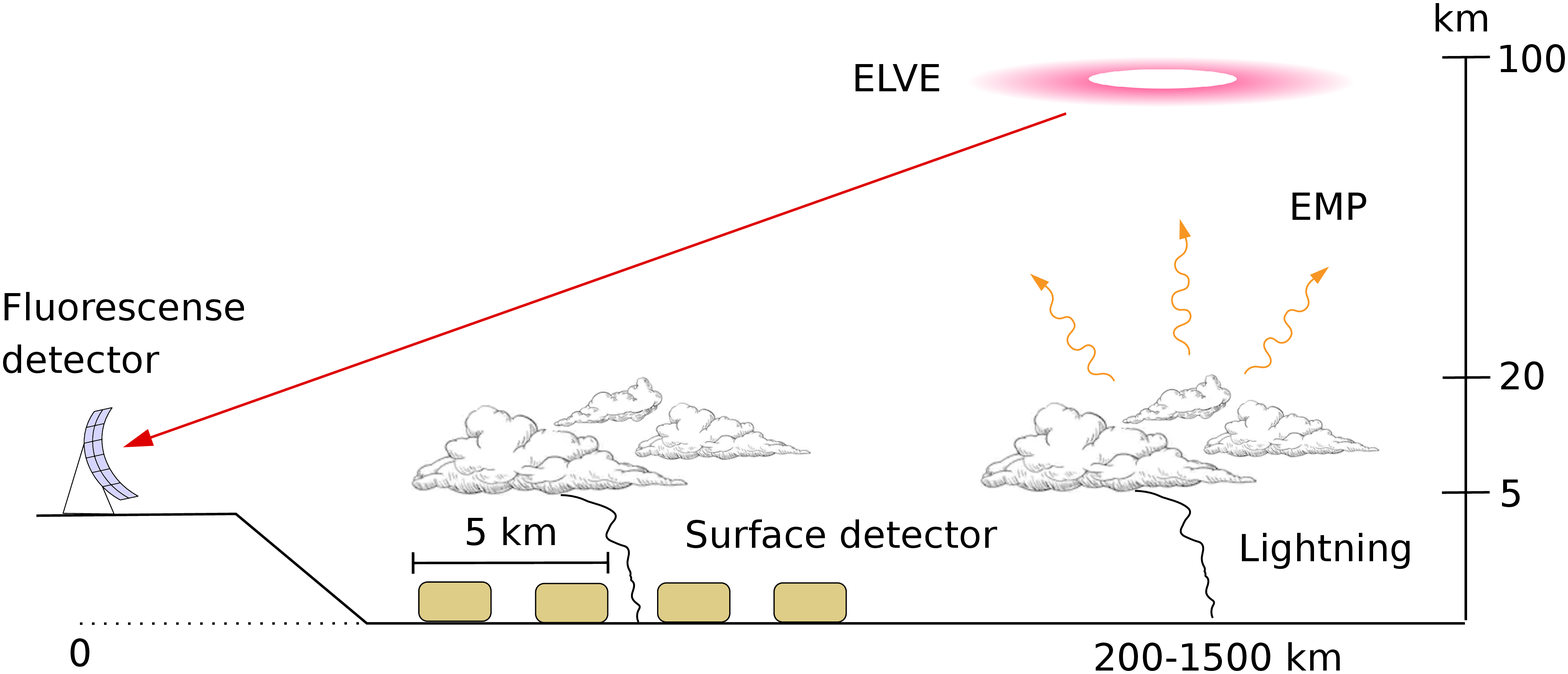}%
\end{wrapfigure}%

\begin{abstract}
During thunderstorms, the atmospheric electric field can increase above hundreds of kV/m, causing an acceleration in the charged particles of secondary cosmic rays. Such an acceleration causes avalanche processes in the atmosphere, enhancing/reducing the particle flux at ground level depending on the strength/polarity of the electric field. We present the design and implementation of a self-triggered and fast-recording lightning monitoring system used to study the transient electric field atmospheric effect on the secondary particle flux above cosmic ray observatories. The acquisition device records lightning electric field at 10\,$\mu$s resolution (during 1.2\,s per event), covering a detection range up to 200\,km ($I_{peak} >$ 100\,kA).
\end{abstract}

\begin{IEEEkeywords}
Lightning, Cosmic Rays, Transient Luminous Event, Particle Detector  
\end{IEEEkeywords}
\end{minipage}}}

\maketitle

\section{Introduction}
\label{sec:introduction}
\IEEEPARstart{C}{osmic Rays} are particles reaching the Earth after crossing interstellar space. High energy cosmic rays (CRs) originate outside the solar system, and most of the low-energy CRs ($<$ 10 GeV) are of solar origin. When CRs enter the terrestrial atmosphere, they collide with the atoms generating Extensive Air Showers (EAS) of secondary particles \cite{Spurio2015}.

The atmospheric electric field rises to hundreds of kV/m during thunderstorm episodes accelerating secondary particles, mainly electrons/positrons, causing Relativistic Runaway Electron Avalanches (RREA) and increasing the particle flux at ground level  \cite{marshall2005observed,dwyer2011low}. The minimum electric field for triggering an RREA is roughly 286\,kV/m \cite{Skeltved2017, colalillo2019}. Some investigations have shown that the atmospheric electric field also modulates the flux of heavier particles such as muons \cite{wang2012effect, alexeenko2002transient}.

The flux variation of the charged particles depends on the polarity of the atmospheric electric field \cite{bartoli2018observation, zhao2019effects}. Negative-charge particles are accelerated towards the ground under a negative electric field (positive charge on the ground and negative charge in the upper atmosphere), enhancing the particle counting at the detector level. On the contrary, under a positive electric field (negative charge on the ground and positive charge in the upper atmosphere), the counting ratio at ground level decreases \cite{dorman2013cosmic}. This behavior can explain the particle flux variation that CR detectors observe during thunderstorms.

There exists another kind of lightning-associated phenomenon detected by fluorescence telescopes in CR experiments \cite{Mussa2012}. Upward accelerated particles from storm clouds interact with nitrogen atoms in the atmosphere causing radiation emission. Sprites, Blue Starters, Glimpses, Blue Jets, Gigantic Jets, ELVES (Emission of Light and Very Low-Frequency perturbations due to Electromagnetic Pulse Sources), and Halos take part in such ultra-fast light emissions and are known as Transient Luminous Events (TLE) \cite{Siingh2015}.

The lack of electric field monitoring networks inside CR observatories limits the understanding of the underlying phenomena responsible for the recorded anomalous events \cite{Colalillo2017, Merenda2019}. Data cross-checking between particle counting rate and transient electric field shall help us clarify the origin of such phenomena.

We outline the development of a Lightning Monitoring Network system (LiMoNet), capable of detecting/recording atmospheric electric field transients during thunderstorm episodes. The low cost and scalability of the device allow us to deploy a temporally synchronized network to permit understanding of the correlations between secondary particle flux and electric field variations above CR observatories.


\section{Methods}

Antennas detect lightning activity using two ways: by measuring variations of the lightning electric field or the magnetic field \cite{Rakov2016}. Cross loop antennas detect lightning magnetic fields, while flat parallel plate ones detect lightning electric fields. The last one has two advantages: the voltage between the plates is proportional to the electric field strength, and deployment of at least three antennas allows us to determine the lightning strike point using the TOA (Time of Arrival) method \cite{Sidik2015}.

\subsection{Parallel-plate antenna}

\begin{figure}[h!]
\begin{center}
\includegraphics[width=0.48\textwidth]{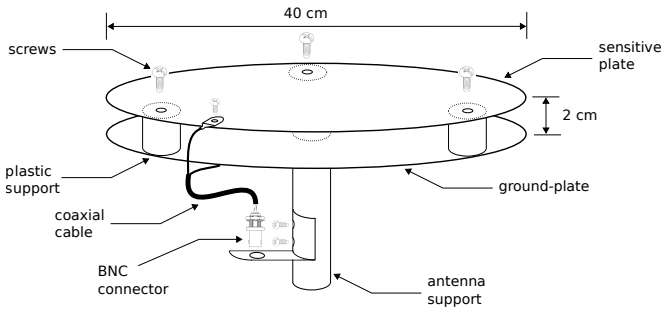}
\caption{Structure of the parallel-plate antenna. Two aluminium plates 40~cm diameter separated by 2~cm constitute the lightning sensor. A coaxial cable connected to a BNC transmits the antenna signal. The bottom plate is grounded.}
\label{antena_rap}
\end{center}
\end{figure}

The parallel-plate antenna consists of two circular aluminium plates 40~cm diameter separated by 2~cm \cite{salgado2020}. The top plate senses the electric field, while the lower one is grounded. The antenna transmits the signal using an RG51 coaxial cable with a BNC connector end. Fig. \ref{antena_rap} displays the antenna structure.

The antenna current due to the induced electric field $E$ is

\begin{equation}
I=\varepsilon_0 A \dfrac{dE}{dt}
\label{eq::antennacurr}
\end{equation}
\noindent where $A$ is the aluminum plate area and $\varepsilon_0$ is the permittivity of free space.

\subsection{Signal integration}

We can deduce from Eq. \ref{eq::antennacurr} that the integration of the antenna signal gives us an estimation of the induced electric field as follows,

\begin{equation}
V = \frac{1}{C} \int I dt = \frac{\varepsilon_0 A E}{C}
\label{eq::antennavol}
\end{equation}
\noindent where $C$ is the antenna capacitance and $V$ the output voltage of the integration stage. The measured electric field is

\begin{equation}
E = \frac{CV}{\varepsilon_0 A}
\label{eq::antennaE}
\end{equation}

\begin{figure}[h!]
\begin{center}
\includegraphics[width=0.5\textwidth]{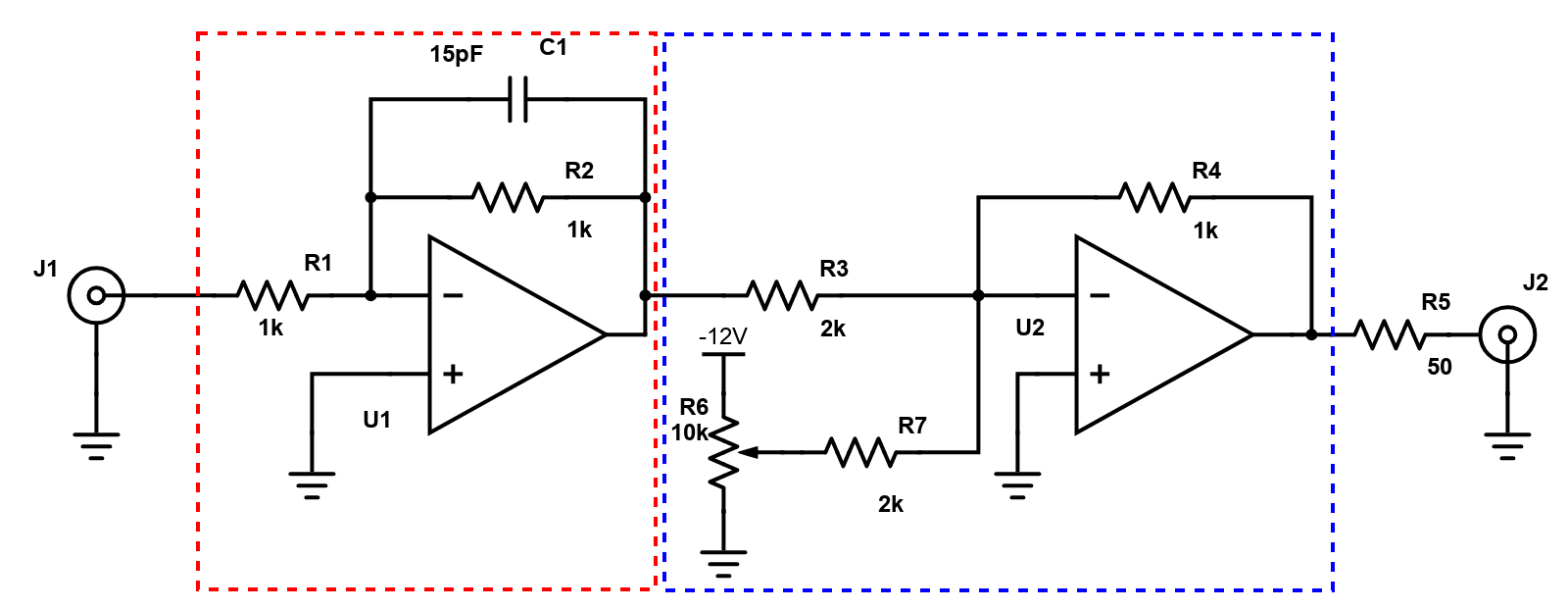}
\caption{Signal conditioning circuit. An active integrator (red-dots) integrates the antenna current signal. An adder (blue-dots) inverts the integrated signal and corrects its baseline.}
\label{fig::integrator}
\end{center}
\end{figure}

An OPA4228 operational amplifier (120\,dB CMRR (Common Mode Rejection Ratio) and 33\, MHz bandwidth) integrates, inverts and increases the base line of the antenna current signal. Fig. \ref{fig::integrator} shows the integrator and baseline circuit.

The circuit's frequency response was simulated with PSpice and tested by varying the input frequency from 100~Hz to 20~MHz. The estimated frequency cut off is at $\sim$10.6~MHz  \cite{salgado2020}.  


\subsection{Signal spectral analysis}

We carried out a spectral analysis of lightning signals previous to the digitization stage. This step helps us to determine parameters at the analogue-to-digital converter sampling frequency.

\begin{figure}[h!]
\begin{center}
\includegraphics[width=0.5\textwidth]{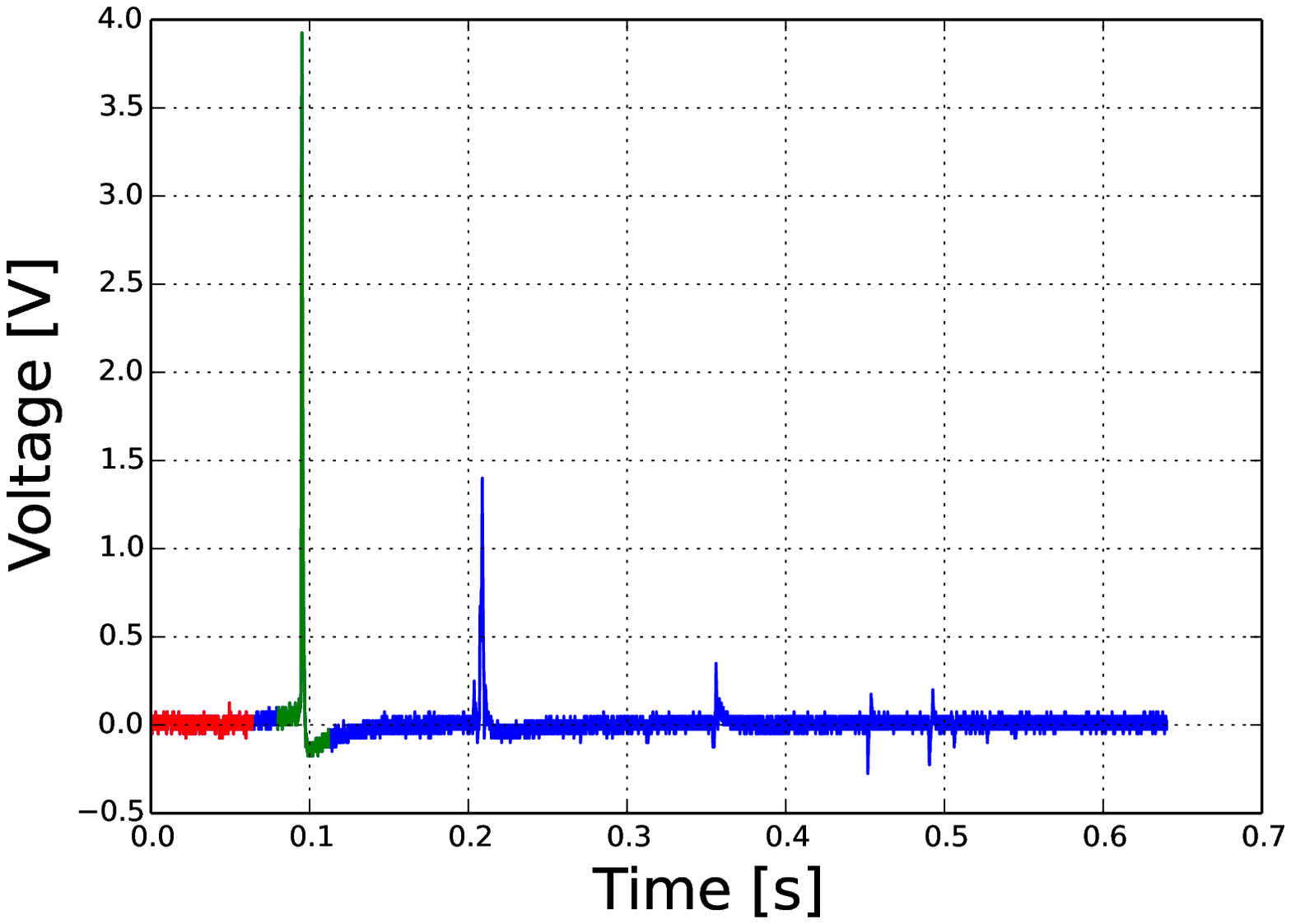},
\includegraphics[width=0.5\textwidth]{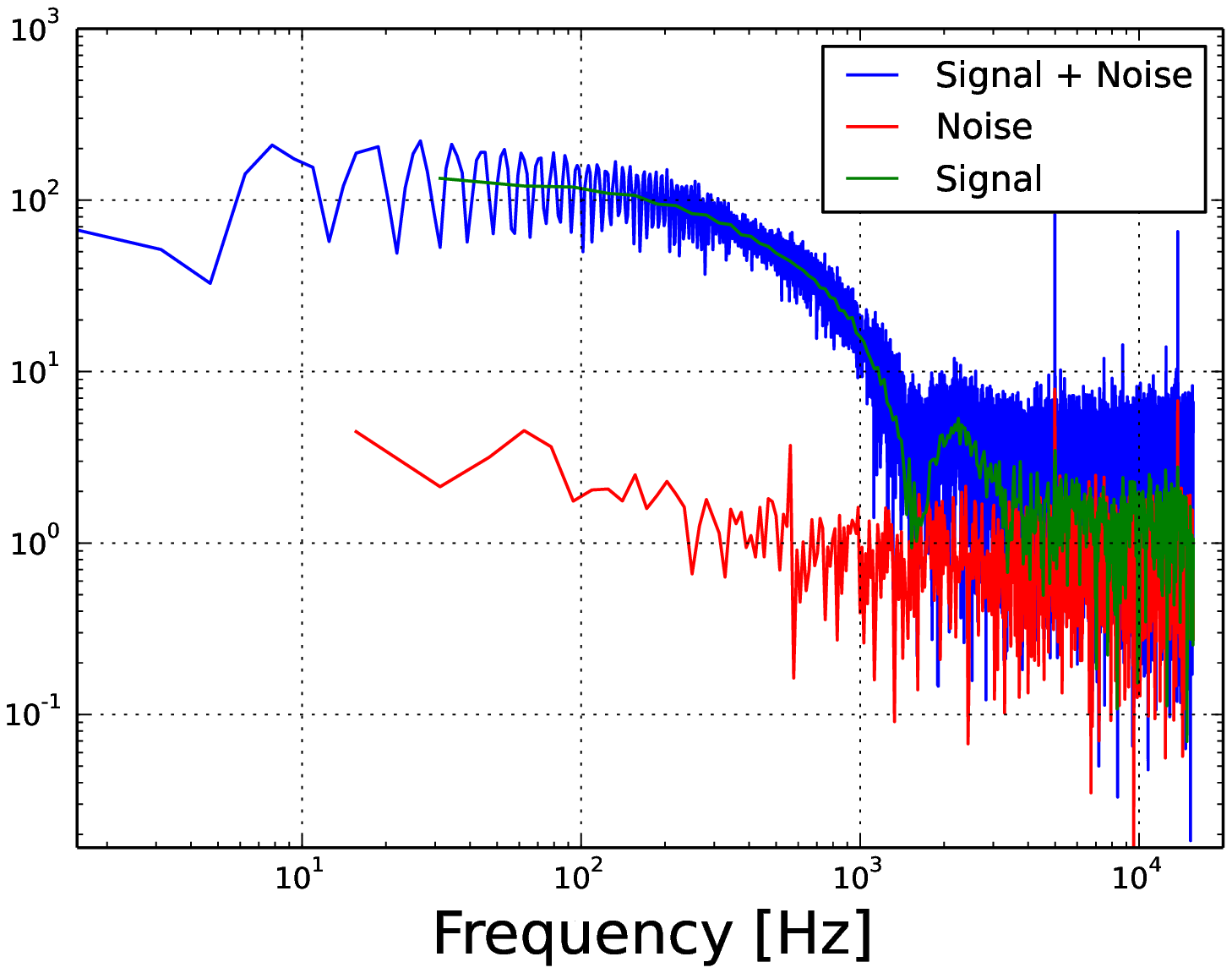}
\caption{(Top) The lightning signal was recorded on March 2, 2020, at Bucaramanga-Colombia. The signal splits into noise (red line) and lightning (green line). (Bottom) Fast Fourier Transform (FFT) of the recorded signal. The frequency components of the discharge (green-line) are below $\sim$4~kHz, and the noise (red-line) extends to $\sim$20~kHz.}
\label{fig::fftsignal}
\end{center}
\end{figure}

A Tektronix TDS 2002B oscilloscope (1\,GS/s) recorded lightning signals directly from the parallel plate antenna. Fig. \ref{fig::fftsignal}-Top shows the 02032020 lightning event recorded at Bucaramanga, Colombia. 

The lightning signal has a first return stroke of $\sim$4\,V (200\,V/m above the antenna) with subsequent discharges at $\Delta t \sim$0.1~s, $\sim$0.25~s, $\sim$0.35~s, and $\sim$0.39~s. The recording window was 0.7~s. 

We characterized the noise (red) and signal (green) spectrum by applying the FFT in two portions of the recorded signal, as shown in Fig. \ref{fig::fftsignal}-Bottom. The main frequency components of the lightning signal are below $\sim$4~kHz, while noise components extend to $\sim$20~kHz.

We set a signal sampling frequency of the acquisition system at $\sim$100\,kHz taking into account the spectral analysis results and that the power spectral density of lightning radiation is in the VLF (3-30 kHz) \cite{Fullekrug2006}.

\subsection{Noise analysis}

\begin{figure}[h!]
\begin{center}
\includegraphics[width=0.45\textwidth]{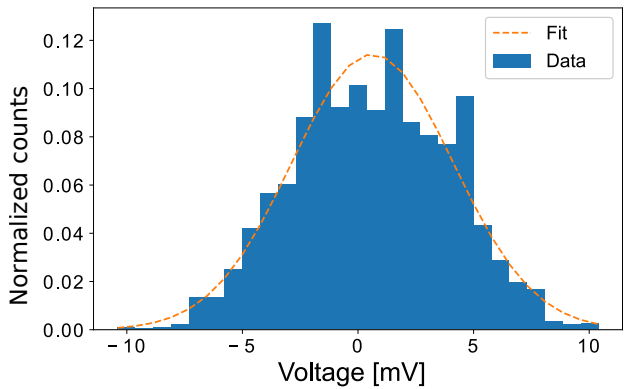}
\caption{Front-end electronics noise distribution. The circuit output varies around the baseline with a standard deviation of 3.49\, mV.}
\label{fig::intnoise}
\end{center}
\end{figure}

We performed a noise analysis of the conditioning circuit to establish the minimum voltage threshold for the lightning triggering stage. We characterized the circuit noise by analyzing its output without a lightning input signal (steady-state). The noise behavior shapes like a normal distribution centered in 0\,mV (white-noise) and a standard deviation $\sim$3.49\,mV as shown in Fig. \ref{fig::intnoise}.

\subsection{Digitization}

A 12-bit ADC AD9235 digitizes the lightning signal at one MHz sampling frequency, and an input range of 2\, V. A Spartan 6 FPGA manages the acquisition process. The FPGA generates a one MHz ADC clock, applies the voltage-over-threshold trigger, and addresses the signal data to 16-Mbit flash memory (SST39VF1602C\footnote{\url{https://www.microchip.com/en-us/product/SST39VF1602C}}), and transmits the data to a Raspberry Pi.

We set a detection threshold of 120\, UADC, considering the noise variability and that the signal undershoots. When the lightning event exceeds the detection threshold, the data acquisition system implements a 1.2\,s time window for recording. The FPGA takes a signal sample from the ADC every ten $\mu$s to reduce the data size and improve system performance. The system also stores 12 signal samples before the trigger to reconstruct the signal shape. Each recorded event has an associated ten ns resolution timestamp and GPS time.

\begin{figure}[h!]
\begin{center}
\includegraphics[width=0.45\textwidth]{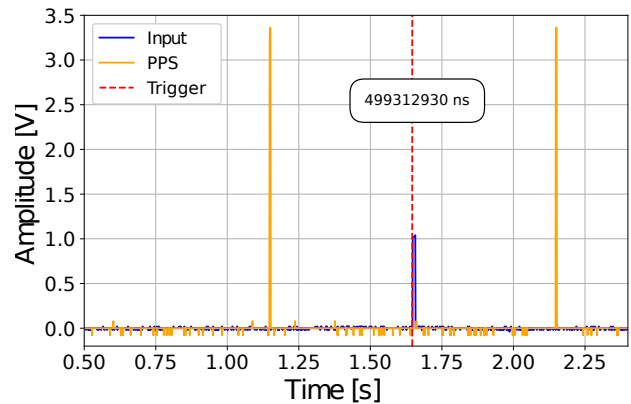}
\caption{The blue line indicates the simulated lightning detection system by a square pulse (50~ms width and 1~V height) from a Tektronix AFG1022 signal generator. The good timestamp (red dashed line) is relative to the PPS signal (yellow line). The occurrence time of the emulated lightning signal was measured $\sim$499~ms. }
\label{fig::pulse}
\end{center}
\end{figure}

We tested the lightning detection hardware by injecting square pulses simulating electrical discharges. Figure \ref{fig::pulse} shows the results using a Tektronix AFG1022 signal generator with 50~ms width and 1~V height. We tuned up the timestamp with the GPS PPS signal (yellow-line). Observe that the emulated lightning signal occurred at $\sim$499~ms from the last PPS, and the red dashed line represents the occurrence time.

\subsection{Monitoring station}

\begin{figure}[h!]
\begin{center}
\includegraphics[width=0.5\textwidth]{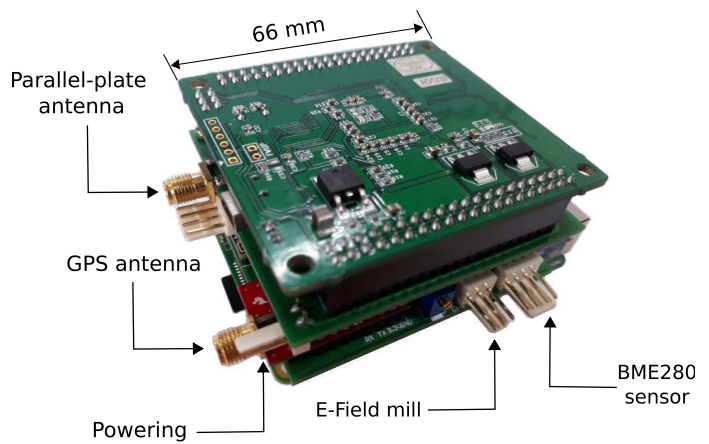}
\caption{Monitoring station hardware. The lightning and environmental readout electronics stack on the Raspberry Pi single computer board. External connectors for powering, weather sensors, GPS antenna, parallel plate antenna, and an E-field mill encircle the station hardware.}
\label{fig::station}
\end{center}
\end{figure}

The monitoring station includes a lightning detection system, an atmospheric electric field mill, and environmental sensors. A single computer board (Raspberry Pi) operates the station running under Raspbian Wheezy. The BME280\footnote{\url{https://www.bosch-sensortec.com/products/environmental-sensors/humidity-sensors-bme280/}} sensor measures temperature, relative humidity, and atmospheric pressure with accuracies of $\pm$1\,$^{\circ}$C, $\pm$3\,$\%$RH, and $\pm$1\,hPa respectively. 

A Venus 638FLPx chip performs GPS positioning with a spatial accuracy of 2.5\,m and time accuracy of 60\,ns, communicating via UART protocol at 9600\,bauds and power with 3.3~V. The positioning data contains Unix timestamp, latitude, N/S indicator, longitude, E/W indicator, and altitude. The time synchronization of the acquisition system depends on the PPS signal. A trilateration algorithm based on the GPS data estimates the lightning strike point \cite{MialdeaFlor2019}. Figure \ref{fig::station} displays the station hardware. 

The four (bottom-up) hardware are the Raspberry Pi 2, the environmental layer (PSOC 5LP, GPS Venus 638FLPx, and BME280), the Spartan 6 FPGA Development Board, and the lightning detection system. In appendixes \ref{Ap:A} and \ref{Ap:B}, we describe the monitoring station mechanics and the data file structure.

\subsection{The Time-of-Arrival method}

There exist several methods for lightning location. The most common include magnetic direction finding (MDF), time of arrival (TOA), and interferometry \cite{Rakov2016}.

Deployment of multiple antennas allows us to locate the lightning strike point and reconstruct the whole 3D lightning channel \cite{Wu2018, Rakov2016}. A single station detects the lightning event but cannot locate the lightning strike point.

LiMoNet uses the TOA approach, operating at VLF (Very Low Frequency) and LF (Low Frequency) range, 3–300 kHz (long baseline systems), to locate the detected lightning events. This frequency band allows LiMoNet to cover a distance range of hundreds to thousands of kilometers. For reconstructing the lightning channel, LiMoNet records the whole waveform in order to perform a waveform correlation method to match pulses at different sites \cite{Wu2018}.

\begin{figure}[h!]
\begin{center}
\includegraphics[width=0.45\textwidth]{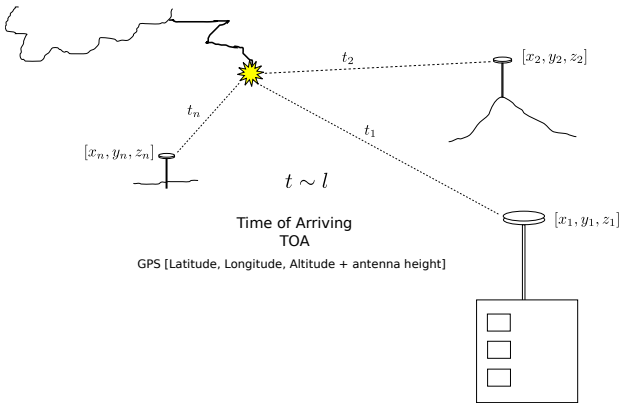}
\caption{Lightning stroke location by using the Time-of-Arrive method. TOA uses the time of arrival differences ($t_1, t_2, \cdots ,t_n$) of the electric field signal at the stations and the station coordinates ($x_i, y_i, z_i$).}
\label{fig::TOA}
\end{center}
\end{figure}

The TOA method locates the lightning strike by comparing the difference in the times of arrival of the signals at the synchronized stations and by knowing the antenna coordinates (latitude$=x_0$, longitude$=y_0$, and altitude + height$=z_0$) \cite{mialdea2019development, Wu2018}, as shown in Figure \ref{fig::TOA}.

\subsection{Strike current model}

Lightning peak current also takes part in the characterization of the phenomena. This parameter helps to establish the possible type of TLE the lightning can trigger. For instance, a cloud-to-ground (CG) lightning with a peak current above 80\,kA generates ELVES \cite{Chang2014}.

\begin{figure}[h!]
\begin{center}
\includegraphics[width=0.45\textwidth]{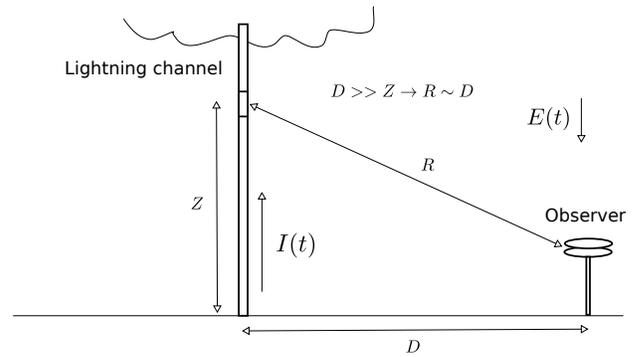}
\caption{Graphical representation of the Transmission Line Model. $Z$ is the lightning channel height, $R$ stands for the distance between the channel and the antenna, $D$ corresponds to the horizontal component of the distance, $I$ be the channel current, and $E$ the measured electric field. If $D>> Z$ then $R \sim D$.}
\label{fig::TL}
\end{center}
\end{figure}

Indirect methods estimate lightning parameters (e.g. current peak) remotely, considering the impossibility of measuring them directly. In the transmission line (TL) model, the current waveform propagates through the atmosphere at a constant speed without changing shape \cite{betz2009, Rakov2016}. The vertical electric field $E (t)$ measured by the antenna at the ground is

\begin{equation}
E (t) = -\frac{\mu_0v}{2 \pi D} I(t-D/c)
\label{eq::efield}
\end{equation}
where $\mu_0 = 1/\epsilon_0 c^2$ the permeability of free space, $v$ the return stroke velocity, $D$ the distance from the lightning to the antenna, $I$ the return stroke current, $t$ the arrival time, and $c$ is the speed of light. The return stroke velocity is assumed to be constant (1-2$\times 10^8$\,m/s) \cite{Rakov2016}. Fig. \ref{fig::TL} shows a graphical representation of the TL model.

The strike current is,
\begin{equation}
I (t) = \frac{2 \pi D}{\mu_0 v} E(t + D/c)
\end{equation}

\section{Results}
\label{Results}
\subsection{Lightning waveform}
In 2021 LiMoNet recorded about 6000 lightning events in the Nort-East zone of Colombia. We divided them into multi-termination (MTF) and single-termination flashes (STF). Single-termination flashes have a single return stroke during the recording window (1.2\,s), and multi-termination flashes have at least two return strokes in the recording window.

\begin{figure}[h!]
\begin{center}
\includegraphics[width=0.48\textwidth]{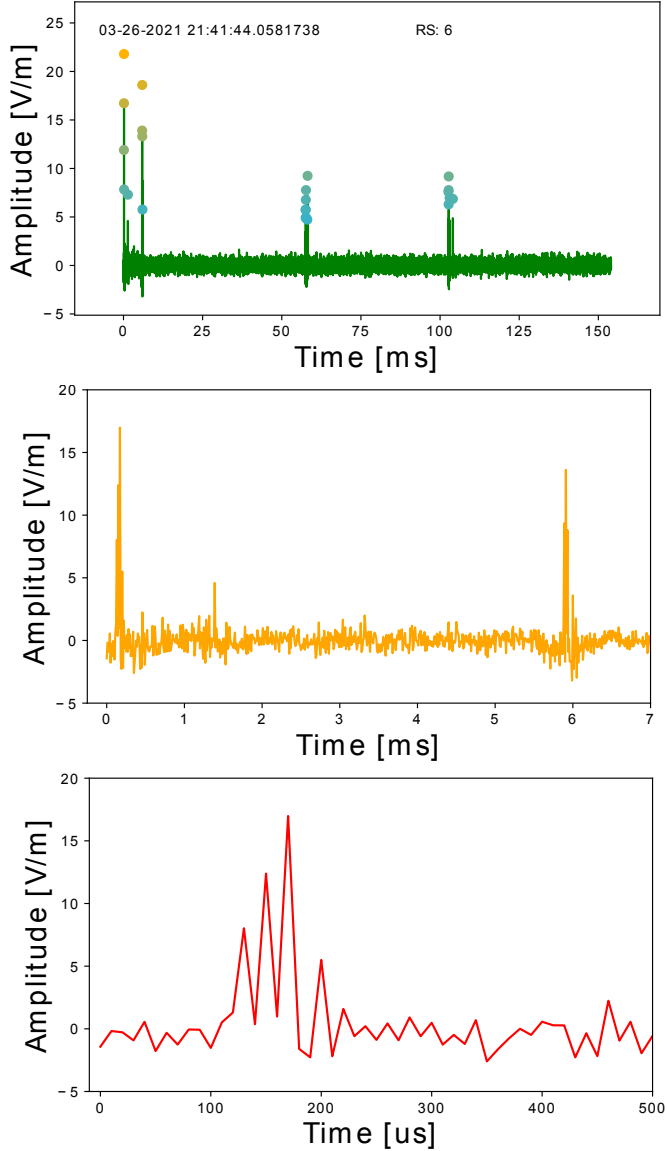}
\caption{(Top) Example of a multi-termination flash with six return strokes. Colored circles represent the intensity of each termination. (Middle) First two return strokes separated by $\sim$6\,ms. (Bottom) Zoom in of the first return stroke. Four terminations in $\sim$100\,$\mu$s compose the initial multi-termination stroke.}
\label{fig::Multi}
\end{center}
\end{figure}

Fig. \ref{fig::Multi} shows the electric field of a multi-termination event recorded by a LiMoNet antenna on March 26, 2021. This flash registered six return strokes in 105\,ms. The first return stroke has four terminations; we call this a multi-termination stroke (MTS). The time interval between the two main strokes is $\sim$6\,ms.

Multi-termination strokes exhibit fast-rising waveforms separated by tens of $\mu$s. The ten \,$\mu$s time resolution of LiMoNet helps reconstruct the waveform shape of multi-termination strokes. MTS could originate multi-ELVES, taking into account the time difference between ELVES peaks ranges 10$\mu$s-100$\mu$s \cite{VsquezRamrez2021}.

After a statistical analysis of the data, we concluded that MTFs mostly have two terminations, and the relative time difference between the first and subsequent strokes was less than 200 ms.

Most MTSs have two terminations and a relative time between them under 50\,$\mu$s. These results correspond with the study carried out by the Fast Antenna Lightning Mapping Array (FALMA) \cite{Gao2019}.

\subsection{Detection sensitivity}

The detection sensitivity of LiMoNet stations depends on the lightning peak current/distance, the detection threshold, and the saturation of the DAQ.

\begin{figure}[h!]
\begin{center}
\includegraphics[width=0.45\textwidth]{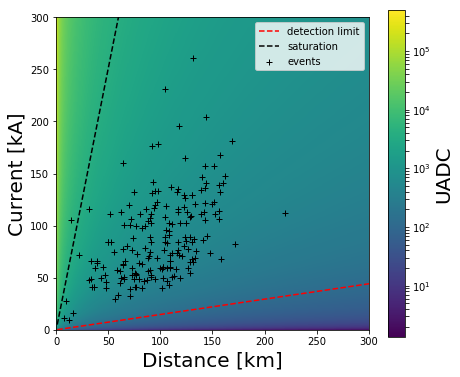}
\caption{LiMoNet antenna sensitivity. The red line determines the minimum detectable peak current threshold (120\,UADC) depending on the lightning strike distance. The black line indicates the data acquisition system's saturation limit (4095\,UADC). Finally, black crosses represent lightning events}
\label{fig::sensitivity}
\end{center}
\end{figure}

The detection threshold establishes the lower boundary of detection sensitivity (120\,UADC). The upper limit of the DAQ (4095\,UADC) sets the saturation limit of the detection system.

We correlated 194 events with a lightning dataset\footnote{\url{https://weather.us/lightning}} to evaluate the sensitivity of detection depending on the distance and peak current of the lightning. The dataset gives us the lightning coordinates and the peak current.

We used the haversine formula for determining the distance $D$ between the antenna (point 1) and the lightning (point 2), given their longitudes and latitudes.

\begin{equation}
\beta = \sin^2 \left( \frac{\phi_2 - \phi_1}{2} \right) + \cos(\phi_1)\cos(\phi_2)\sin^2 \left( \frac{\lambda_2-\lambda_1}{2} \right)
\label{eq::distance}
\end{equation}

\begin{equation}
D=2r \arcsin \sqrt{\beta}
\label{eq::distance}
\end{equation}
where $\phi_1, \phi_2$ are the latitude of point 1 and latitude of point 2, and $\lambda_1, \lambda_2$ are the longitude of point 1 and longitude of point 2.

Fig. \ref{fig::sensitivity} shows the LiMoNet sensitivity evaluation result. The lower limit (red-line) determines the minimum detectable peak current depending on the distance of the lightning. The upper limit (black-line) determines the maximum detectable peak current without saturating the recording system.

\subsection{Detection range}

\begin{figure}[h!]
\begin{center}
\includegraphics[width=0.45\textwidth]{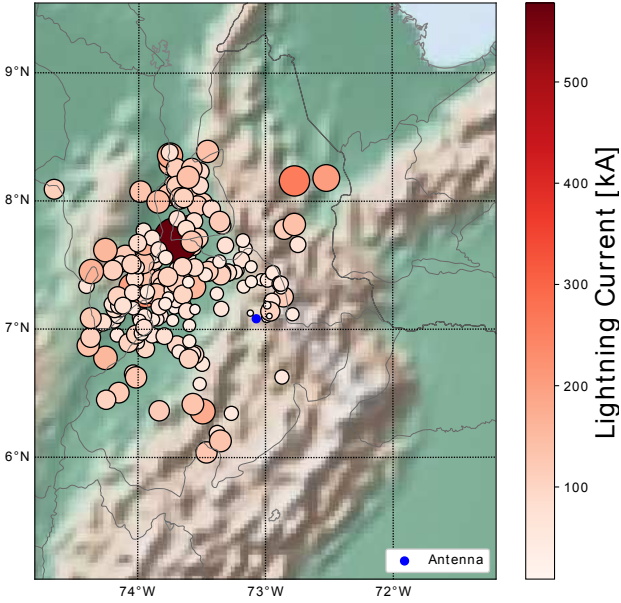}
\caption{Example of 194 events (orange-red circles) detected by a LiMoNet antenna (blue circle). The most decisive event (red-circle) detected had a peak current of -578\,kA at 103\,km. The highest lightning activity region is called the Mid Magdalena zone.}
\label{fig::map}
\end{center}
\end{figure}

We also evaluated the detection distance range of the LiMoNet stations. Fig. \ref{fig::map} displays 194 detected events. The orange-red circles indicate the current peak magnitude, and the blue circle the LiMoNet antenna coordinates.

The most powerful lightning strike detected by LiMoNet (10/21/2021 01:32:10) had a peak current of -578\,kA and a distance of 103\,km from the antenna. We observed a hot spot region (Magdalena Medio) of high lightning activity shaped by two mountain ranges (the Cordillera Central and the Cordillera Oriental). 

The furthest lightning was at a distance of 209\,km from the antenna with a peak current of -112\,kA. The estimation of the LiMoNet antenna range and detection sensitivity allow us to design the distribution of the monitoring network for a given area.

\subsection{Current estimation}

\begin{figure}[h!]
\begin{center}
\includegraphics[width=0.45\textwidth]{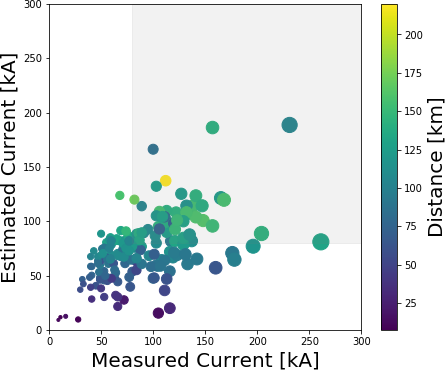}
\caption{Estimated TL model vs measured peak current. The grey zone indicates lightning with a peak current above 80\,kA (lower limit for ELVES production). The color bar indicates the strike distance from the antenna.}
\label{fig::current}
\end{center}
\end{figure}

We estimated the lightning peak current using the TL model and compared the results with the peak current measured by the reference dataset. Fig. \ref{fig::current} shows the estimated/measured lightning peak current and the strike distance. We found a linear agreement of the TL model peak current estimation and measurements with a ratio $a$ (slope) $\sim$0.92. The parameter $a$ shall enhance the TL model peak current estimation.

\begin{equation}
I (t) = \frac{1}{a} \frac{2 \pi D}{\mu_0 v} E(t + D/c)
\label{eq::current}
\end{equation}

The geographical morphology of the studied zone determines the data dispersion of the scatter plot. A more accurate model could be implemented by taking into account attenuation by geological structures, but this approach is not addressed in this study.

\vspace{3cm}

\section{CONCLUSION}

Several studies have demonstrated a direct relationship between high-intensity variations of the atmospheric electric field and the flux of charged particles at ground level caused by RREA or TLE. In this note, we present a lightning monitoring station designed for fulfilling timing and triggering requirements to correlate lightning events with counting rate measurements of particle/fluorescence detectors in cosmic ray observatories.

The LiMoNet DAQ system reconstructs the MTS waveform based on its 10\,$\mu$s resolution. The LF/VLF band operation allows LiMoNet to span a detection range of hundreds of kilometers. LiMoNet continuously recorded events during 2021, detecting $\sim$6000 lightning strikes in a radius of $\sim$209\,km.

The LiMoNet also recorded one of the most powerful lightning strikes of 2021 in Colombia (10/21/2021 01:32:10), a cloud-to-ground discharge of -578\,kA in the Mid Magdalena zone. These extreme strikes have a high possibility of being associated with TLEs.

\section*{ACKNOWLEDGMENT}
The authors recognise the financial support of Vicerrector\'{i}a Investigaci\'on y Extensi\'on Universidad Industrial de Santander and its permanent sponsorship. We are also very thankful to the Pierre Auger Collaboration for their continuous inspiration and support.

\appendices

\section{Mechanical structure}
\label{Ap:A}

A waterproof plastic enclosure (IP65 standard) contains the station hardware. The box has a transparent cover to monitor the station status LEDs. Four plastic screws attach the enclosure cover with the body separated by a rubber seal.

Two case external SMA connectors input the GPS and parallel plate antenna signals. A 4-pin XLR connects the electric field mill (signal, phase, and GND). A second 4-pin XLR connector inputs the environmental sensor BME280 (SDA, SCL, 3.3~V, and GND), and a 7-pin XLR connector powers the station (5~V and GND). All the connectors are on the bottom side of the plastic enclosure.

\section{Data files}
\label{Ap:B}

The monitoring station runs concurrently two acquisition python codes. The first records the atmospheric electric field and environmental data, and the second stores the lightning data.

The environmental acquisition code hourly creates a data file named \textbf{Datos\_YYYY\_MM} \textbf{\_DD\_HH.dat} (YYYY-year, MM-month, DD-day, and HH-hour), taking into account the UTC data from the GPS. The environmental data file starts with the station metadata (Station name, code version, and geographical location) labelled with \textbf{\#} character. Sensors data contain UTC, atmospheric electric field, temperature, pressure, and humidity. The PPS signal stars every new data row acquisition.

The lightning detection code starts the acquisition when an interruption signal from the FPGA trigger system detects a new event. This methodology avoids memory waste typical of continuous recording at high sampling frequencies. A file named \textbf{Lighting\_YYYY\_MM} \textbf{\_DD\_HH\_mm.dat} (mm-minute) stores the lightning waveform (120000 samples - 1.2~s) and the nano-second time stamp. The file metadata contains the code version, station geo-position (latitude and longitude), vertical resolution (2.048~mV/UADC), sampling period (10~$\mu$s), and event time (seconds).

\begin{lstlisting}[language=bash, caption=Metadata]
# # LiMoNet 
# # This is a LiMoNet raw data file 
# ID LM02307a21
# Version 2.0
# ORCID 0000-0002-9861-1023
# Latitude 0708.3503 N
# Longitude 07307.2759 W
# Altitude 1019.3 m
# GPS  Active
# Model Venus638FLPx
# H 480 cm : Antenna height 
# D 2 cm : Plate separation 
# R 488 uV/UADC : ADC resolution 
# C 36884594 : Clock counter (10 ns step) 
# S 1633523090 : Unix time 
# F 10 us : Sampling frequency (1/F) 
# # Time[s] UADC : Data columns
\end{lstlisting}

\bibliographystyle{IEEEtran}
\bibliography{LiMoNet.bib}

\begin{IEEEbiography}[{\includegraphics[width=1in,height=1.25in,clip,keepaspectratio]{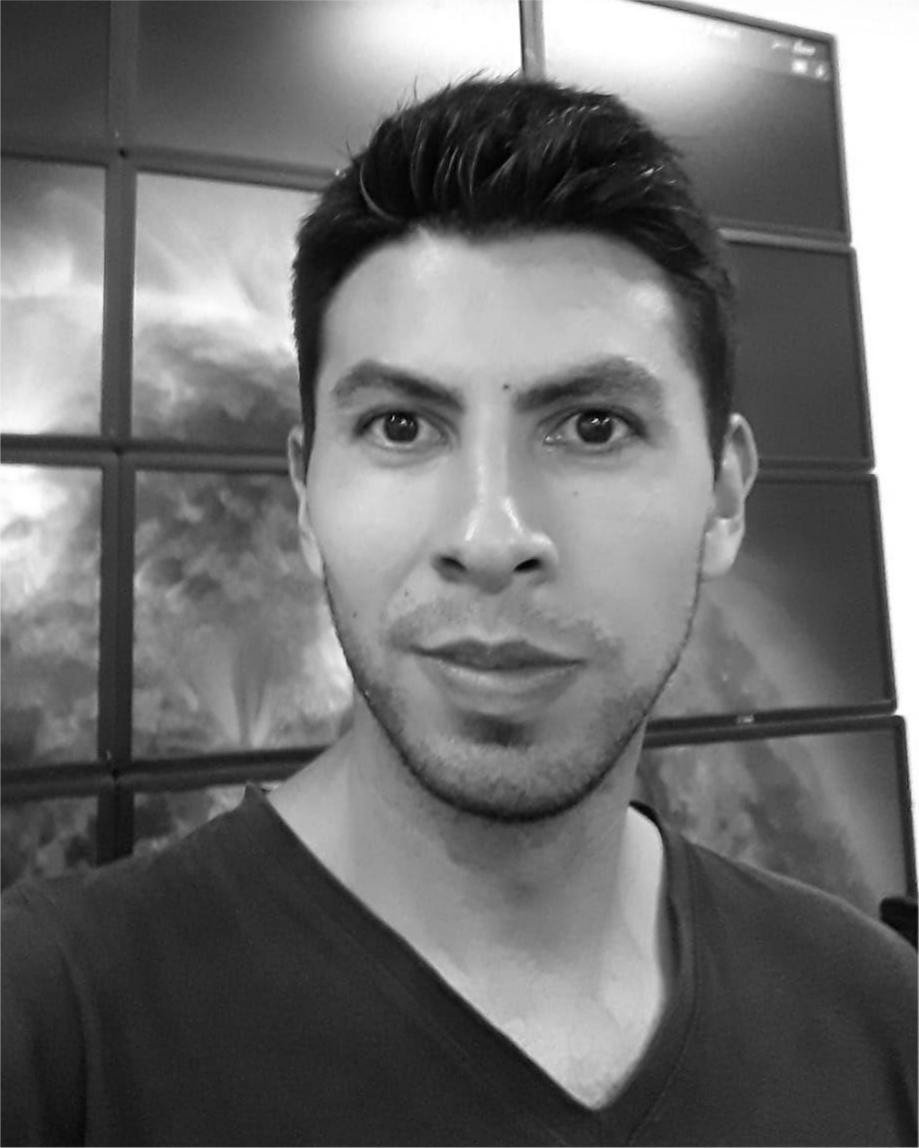}}]{J. Pe\~na-Rodr\'iguez} received the B.S. degree in electronics engineering from the Universidad de Pamplona in 2011 and the M.S. degree in electronics engineering from the Universidad Industrial de Santander in 2016. He holds a Ph.D. degree in physics from the Universidad Industrial de Santander.

From 2014 to 2021, he was a Research Assistant with the Grupo GIRG in the particle detector instrumentation area applied to space weather and muography. Currently, he is Associated Professor at Universidad Santo Tomás. His research interests include particle physics and applications, machine learning, lightning detection, and weather monitoring. He takes part in the LAGO collaboration and the Pierre Auger Observatory.

\end{IEEEbiography}

\begin{IEEEbiography}[{\includegraphics[width=1in,height=1.25in,clip,keepaspectratio]{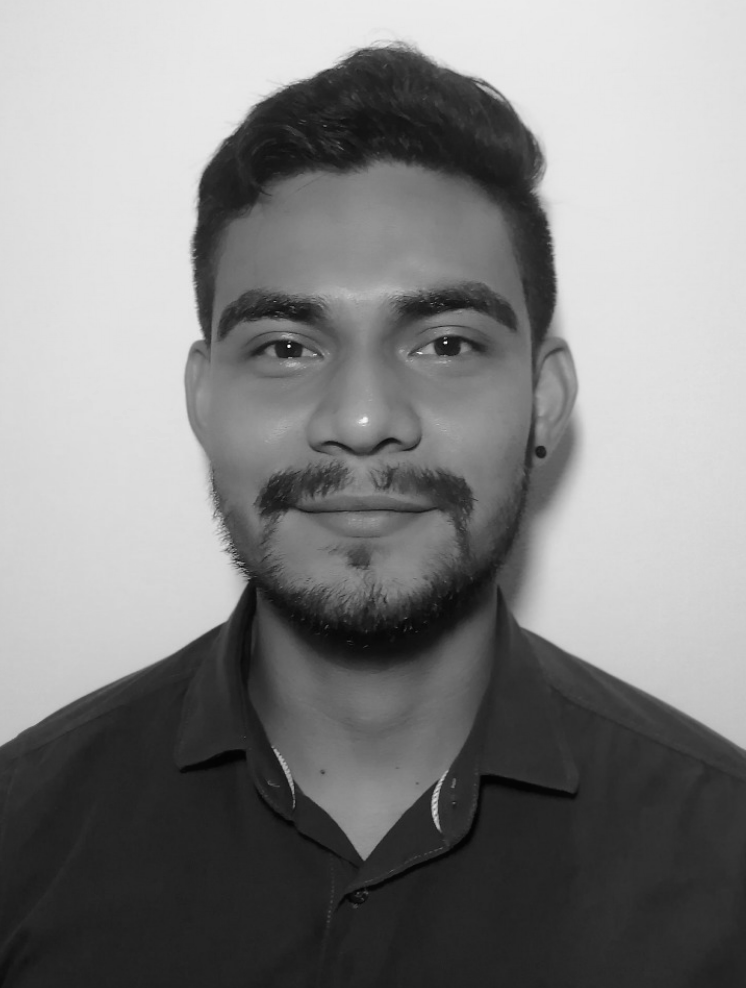}}]{P. Salgado-Meza} received the B.S. degree in electronics engineering from the Universidad Industrial de Santander in 2020. He worked in the design and assembly of the lightning monitoring station LiMoNet. He is studying a M.S. degree in Telecommunications at Universidad Industrial de Santander.

\end{IEEEbiography}

\begin{IEEEbiography}[{\includegraphics[width=1in,height=1.25in,clip,keepaspectratio]{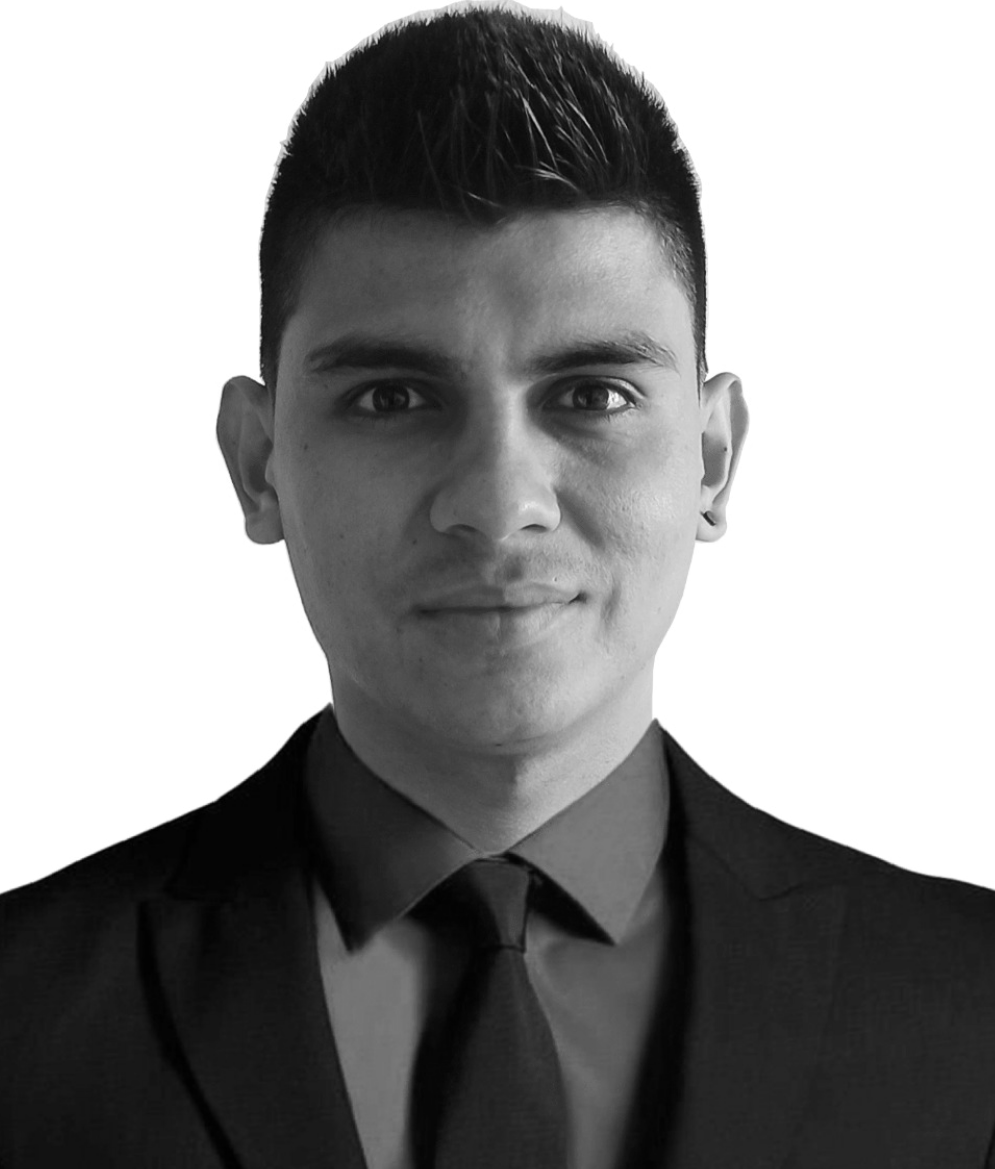}}]{L. Flórez-Villegas} received the B.S. degree in electronics engineering from the Universidad Industrial de Santander in 2020. He worked in the design and assembly of the lightning monitoring station LiMoNet. He is interested in control systems and industrial automation.

\end{IEEEbiography}

\begin{IEEEbiography}[{\includegraphics[width=1in,height=1.25in,clip,keepaspectratio]{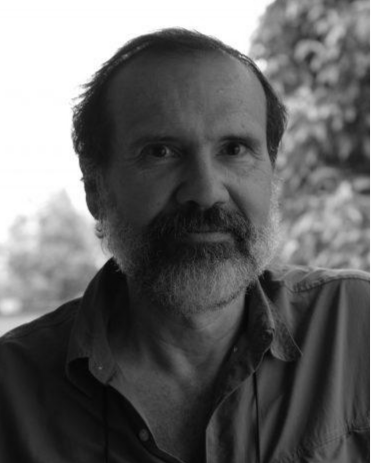}}]{L. A. N\'u\~nez} is a Senior Professor at Universidad Industrial de Santander, Bucaramanga-Colombia since 2010. From 1979 to 2009 he was a Former Senior Professor at Universidad de Los Andes, Mérida-Venezuela. Director of Information Technologies at the Universidad de los Andes from 1995 to 2009, and Director of the Venezuelan National Center for Scientific Computing (CeCalCULA) from 1999 to 2009. He has been involved in several European cooperation projects with Latin America.

He presently is a member of the Latin American Giant Observatory Collaboration (LAGO), the Pierre Auger Observatory Collaboration, and the Cooperación Latinoamericana de Redes Avanzadas (RedCLARA). His areas of interest are astroparticle physics, relativistic astrophysics, and information technologies.
\end{IEEEbiography}

\end{document}